\documentstyle[sprocl]{article}

\input{epsfig.sty}

\def\Journal#1#2#3#4{{#1} {\bf #2}, #3 (#4)}

\def\PTP{\em Prog. Theor. Phys.}
\def\PTPS{\em Prog. Theor. Phys. Suppl.}
\def\soken{\em Soryushiron Kenkyu (Kyoto)}
\def\NPA{{\em Nucl. Phys.} A}
\def\PLB{{\em Phys. Lett.}  B}
\def\PR{\em Phys. Rev.}
\def\PPNP{\em Prog. Part. Nucl. Phys.}
\def\PRC{{\em Phys. Rev.} C}
\def\ZPA{{\em Z. Phys.} A}
\def\AP{\em Ann. Phys.}
\def\ibid{\it ibid.}
\def\JPG{{\em J. Phys.} G}

\begin{document}

\title{RELATIVISTIC STUDY OF MEDIUM POLARIZATION EFFECTS
ON SUPERFLUIDITY IN NUCLEAR MATTER}

\author{MASAYUKI MATSUZAKI}

\address{Department of Physics, Fukuoka University of Education,\\
Munakata, Fukuoka 811-41, Japan\\
and\\
Physik-Department der Technischen Universit\"at M\"unchen,\\
D-85748, Garching, Germany\\
E-mail: matsuza@fukuoka-edu.ac.jp}

\author{PETER RING}

\address{Physik-Department der Technischen Universit\"at M\"unchen,\\
D-85748, Garching, Germany\\
E-mail: ring@physik.tu-muenchen.de}

\maketitle\abstracts{
Effects of the Fermi-sea polarization on the superfluidity in nuclear matter
are studied in a framework of quantum hadrodynamics.  The particle-hole
polarization due to $\sigma$ and $\omega$ mesons enhances the peak value of
the pairing gap contrary to the previous non-relativistic studies.}

\section{Introduction}

  Pairing correlation in nuclear matter has long been studied mainly in neutron
matter from a view point of neutron-star physics such as cooling
rates~\cite{one}.  In addition, pairing in
nuclear matter with finite $Z/N$ ratio is also becoming of interest as a
basic information for the structure theory of finite nuclei, since recent
development of RI-beam experiments makes it possible to study $N \simeq Z$
medium-heavy nuclei and neutron-rich light nuclei.

  At present, there are two ways to describe the fundamental properties such as
the saturation property of the finite-density nuclear many-body system; the
non-relativistic and the relativistic models. They are understood as describing
observed properties almost equally.  Among them, we here adopt the latter
because of its relative proximity to the underlying field theory.  The origin
of quantum hadrodynamics (QHD) can be traced back to Duerr's relativistic
nuclear model~\cite{two} which reformulated a non-relativistic field
theoretical model of Johnson and Teller~\cite{jt}.  Since Chin and Walecka
succeeded in reproducing the saturation property of symmetric nuclear matter
within the mean-field approximation~\cite{cw},
QHD has not only been evolving beyond the mean-field approximation as a
many-body theory but also been enlarging its objects as a nuclear structure
model as infinite matter $\rightarrow$ spherical $\rightarrow$ deformed
$\rightarrow$ rotating nuclei~\cite{rev}.  These successes indicate that the
particle-hole interaction in QHD is realistic.  In contrast,  relativistic
nuclear structure calculations with pairing done so far have been using
particle-particle interactions
borrowed from non-relativistic models and therefore the particle-particle
channel in QHD has not been studied well even in infinite matter.  Aside from
practical successes, this situation is unsatisfactory theoretically.
Therefore, in this talk, we present an effort to derive an in-medium
particle-particle interaction which is consistent with the relativistic mean
field although only infinite matter can be discussed at the present stage.

  Up to now, there have been a lot of non-relativistic studies of the pairing
in nuclear matter.  As for the particle-particle interaction entering into
the gap equation, some authors adopted bare interactions whereas others
adopted renormalized ones such as $G$-matrices or with Jastrow correlations.
Although, in the medium, renormalized interactions should be used intuitively,
following reasons support the use of bare interactions: 1) The Green's
function formalism leads to the sum of the irreducible
diagrams~\cite{three}$^,$\cite{four}, and its
lowest order is the bare interaction.  2) The gap equation itself implies the
short-range correlation~\cite{five}$^,$\cite{six}$^,$\cite{one}$^,$\cite{four}.
In general, medium renormalizations are expected to enhance the gap by
reducing the short-range repulsion.  Anyway, as a next step, polarization
diagrams should be considered.  One of the formulations which incorporate both
the ladder and the ring diagrams consistently is Babu-Brown's~\cite{seven}.
Some works~\cite{eight}$^,$\cite{nine} were
done based on this.  Among them, ref.\ref{refeight} considered momentums around
the Fermi surface only, while ref.\ref{refnine} solved the gap equation in the
full momentum range.  The other~\cite{ten} is based on the correlated basis
function method.  All these works concluded that the inclusion of the
polarization reduced the pairing gap.

  On the other hand, there are a few relativistic calculations of the
pairing gap.  The first one was done by Kucharek and Ring~\cite{eleven}.
They adopted the one-boson-exchange (OBE) interaction with the ordinary QHD
parameters
(both the linear and the non-linear sets), which gave the saturation, under
the no-sea approximation.  The resulting maximum pairing gap given by the
full-range gap equation was about three times larger than the accepted
values in the non-relativistic calculations.  It should be noted, however,
that whether the full-range calculation with effective forces is adequate
or not is not trivial as discussed in ref.\ref{refeleven}.  In this respect,
Matera et al. discussed a prescription to introduce a cut-off of the
high-momentum region~\cite{thirteen}.  Although Guimar\~aes et al.
reported that the inclusion of the negative-energy states improved the
result~\cite{twelve}, here we concentrate on the effects of the Fermi-sea
polarization to differentiate pure relativistic effects from those common to
non-relativistic models.

\section{Outline of formulation}

  As described in ref.\ref{refeleven}, meson fields also have to be treated
quantum
mechanically to incorporate the pairing field via the anomalous (Gorkov)
Green's functions.  The resulting Dirac-HFB equation reduces to the ordinary
BCS equation in the infinite matter case.  Therefore, the actual task is
to solve the coupled equations:
\begin{equation}
\begin{array}{rcl}
  M^\ast &=& M-{{g_s^2}\over{m_s^2}}{\gamma\over{2\pi^2}}
   \int_0^\Lambda {M^\ast\over\sqrt{\vec{k}^2+M^{\ast\,2}}}v^2(k)k^2dk,\\
  \Delta(p) &=& -{1\over{8\pi^2}}
 \int_0^\Lambda v_{\rm pp}(p,k)
         {\Delta(k)\over\sqrt{(e_k-e_{k_{\rm F}})^2+\Delta^2(k)}}k^2dk,\\
  v^2(k) &=& {1\over2}(1-
   {{e_k-e_{k_{\rm F}}}\over\sqrt{(e_k-e_{k_{\rm F}})^2+\Delta^2(k)}}),\\
  e_k &=& \sqrt{\vec{k}^2+M^{\ast\,2}}+g_vV_0 ,\\
\end{array}\label{eq:eqa}
\end{equation}
where $v_{\rm pp}(p,k)$ is an angle-integrated, anti-symmetrized matrix element
of the adopted particle-particle interaction, $\Lambda$ is a cut-off momentum,
and $\gamma=$4 (2) for symmetric nuclear matter (neutron matter).  A natural
choice of $v_{\rm pp}$ is the OBE without form factors, with the coupling
constants
and meson masses which are consistent with the mean field.  Note that the
mesons whose ground-state expectation values are zero (such as $\pi$) can
also contribute here.  Since it was shown in ref.\ref{refeleven} that the
pairing field was determined mainly by $\sigma$ and $\omega$, however, we take
into account only these two in the present polarization calculation.

  Here we examine the Fermi-sea polarization diagrams added to the OBE terms
mentioned above.  The polarization term due to $\sigma$, for example, is given  by
\begin{equation}
  v_{\rm pp}^{({\rm pol})}(p,k)
  =({g_s\over{q^2-m_s^2}})^2\Pi^s(q)(\bar uu)_1
                                    (\bar{\tilde u}\tilde u)_2,
\label{eq:eqb}
\end{equation}
with a 4-momentum $q=k-p$, and $u$ stands for the nucleon spinor with tildes
indicating time-reversal.  Similar terms due to
$\omega$ and the $\sigma$-$\omega$ mixing are also included.  As for the
polarization insertion $\Pi(q)$, we examine the ordinary Feynman-density
decomposition and the particle-hole-antiparticle decomposition~\cite{forteen}
in order to single out the pure particle-hole effects which compare with those
in the non-relativistic
calculations.  In the actual calculation, an instantaneous approximation
($q^0=0$) was adopted as in refs.\ref{refeleven} and \ref{refthirteen}.  Note
that an spin sum, an angle integration and an anti-symmetrization are
necessary to obtain the matrix elements in the $^1S_0$ gap equation in
(\ref{eq:eqa}).  The following results are given by a one-ring
calculation.  A preliminary RPA calculation shows that the collectivity does
not change the essential feature.  

\section{Results and discussion}

\begin{figure}[h]
\begin{center}
\epsfig{figure=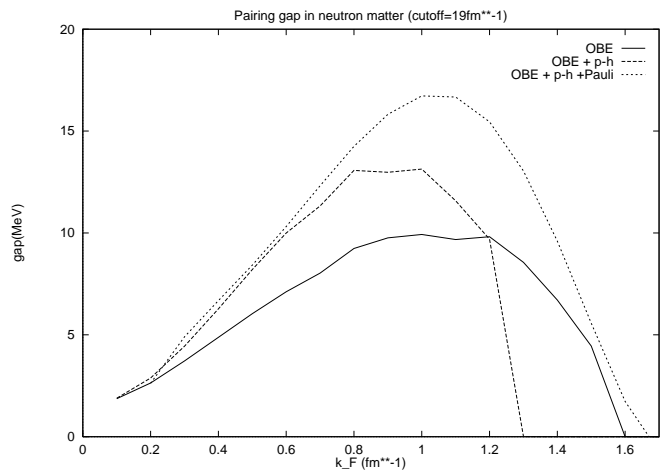,width=9cm}
\end{center}
\caption{Pairing gap in neutron matter is shown as a function of the Fermi
momentum.
The solid, the dashed and the dotted lines indicate the OBE,
the OBE + p-h polarization and the OBE + p-h polarization + Pauli blocking
cases, respectively.  The second and the third ones were given by the p-h-a
and the Feynman-density decompositions, respectively.  Parameters used are
$\Lambda=$19 fm$^{-1}$, $m_s=$520MeV, $m_v=$783MeV, $g_s=$9.051 and
$g_v=$11.672.}
\label{fig:figa}
\end{figure}

\begin{figure}[p]
\begin{center}
\epsfig{figure=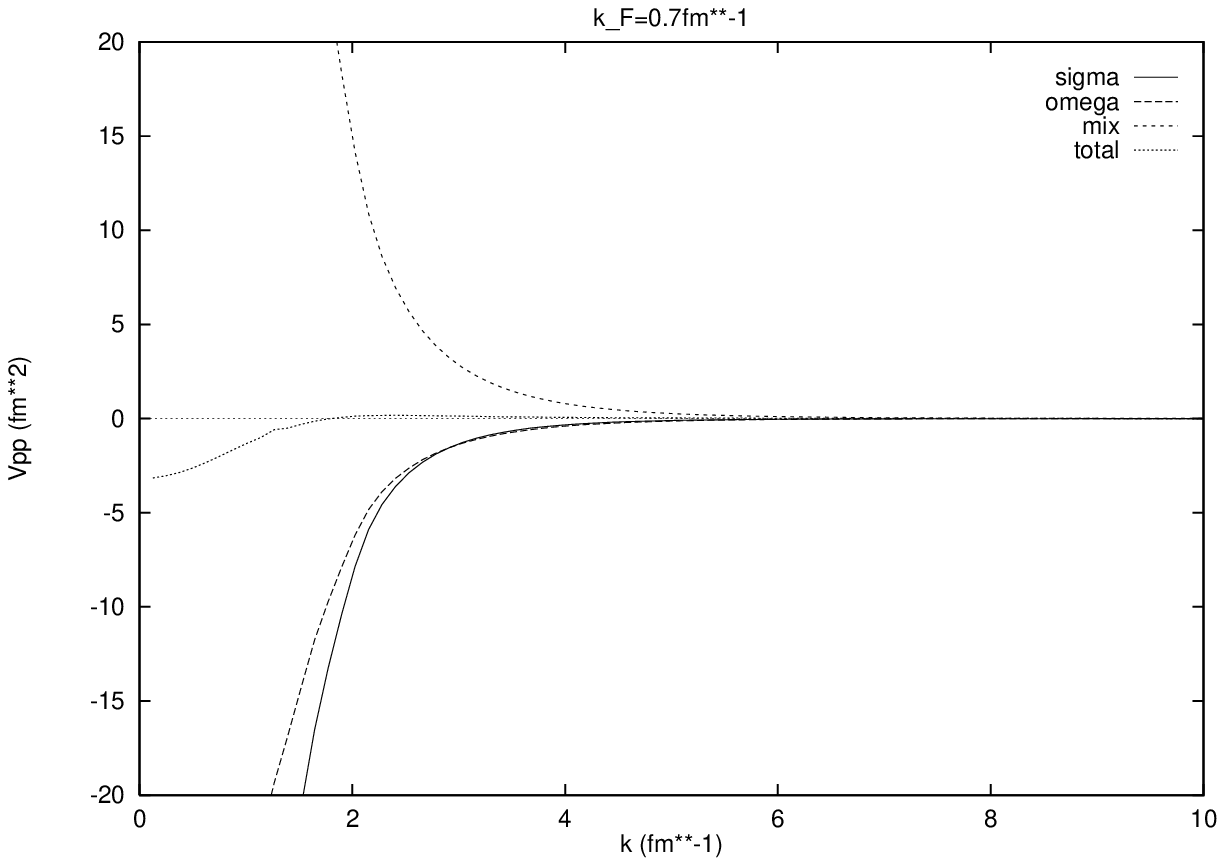,width=7cm}
\end{center}
\begin{center}
\epsfig{figure=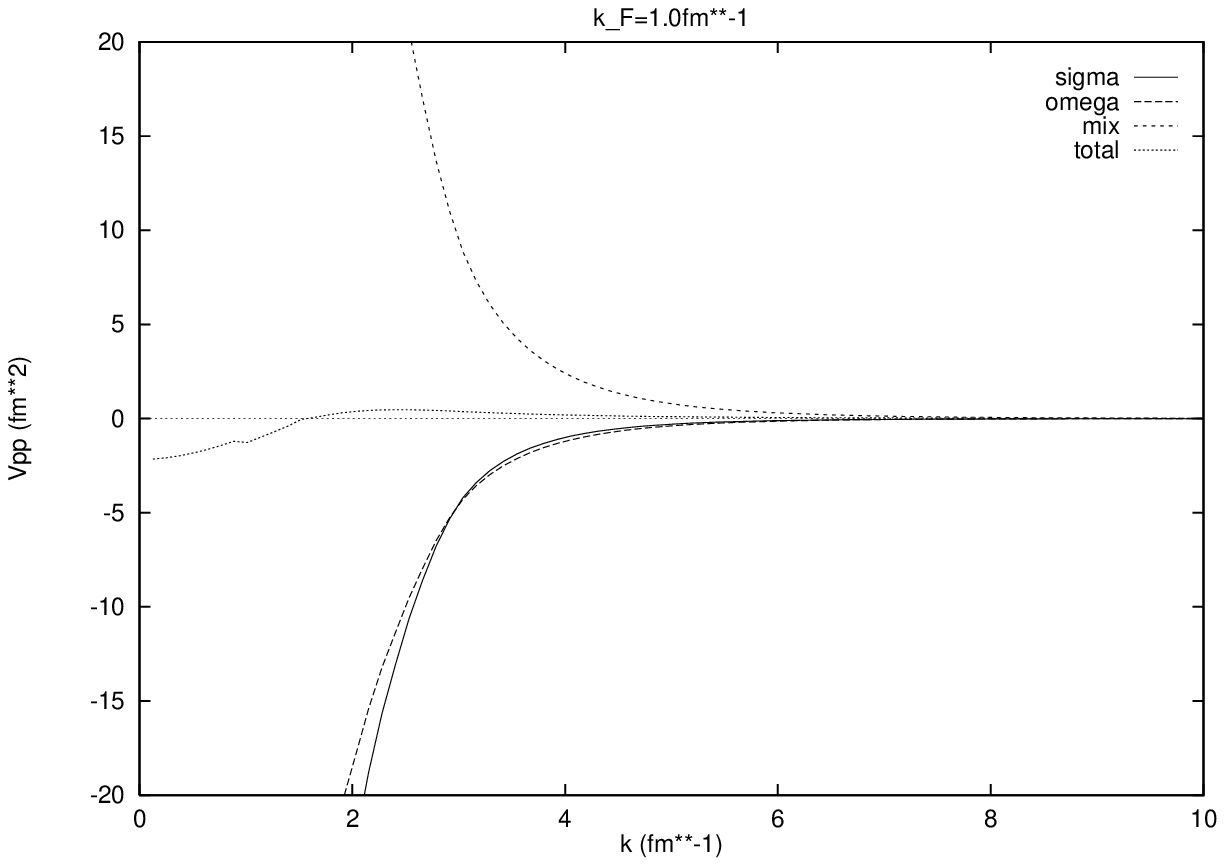,width=7cm}
\end{center}
\begin{center}
\epsfig{figure=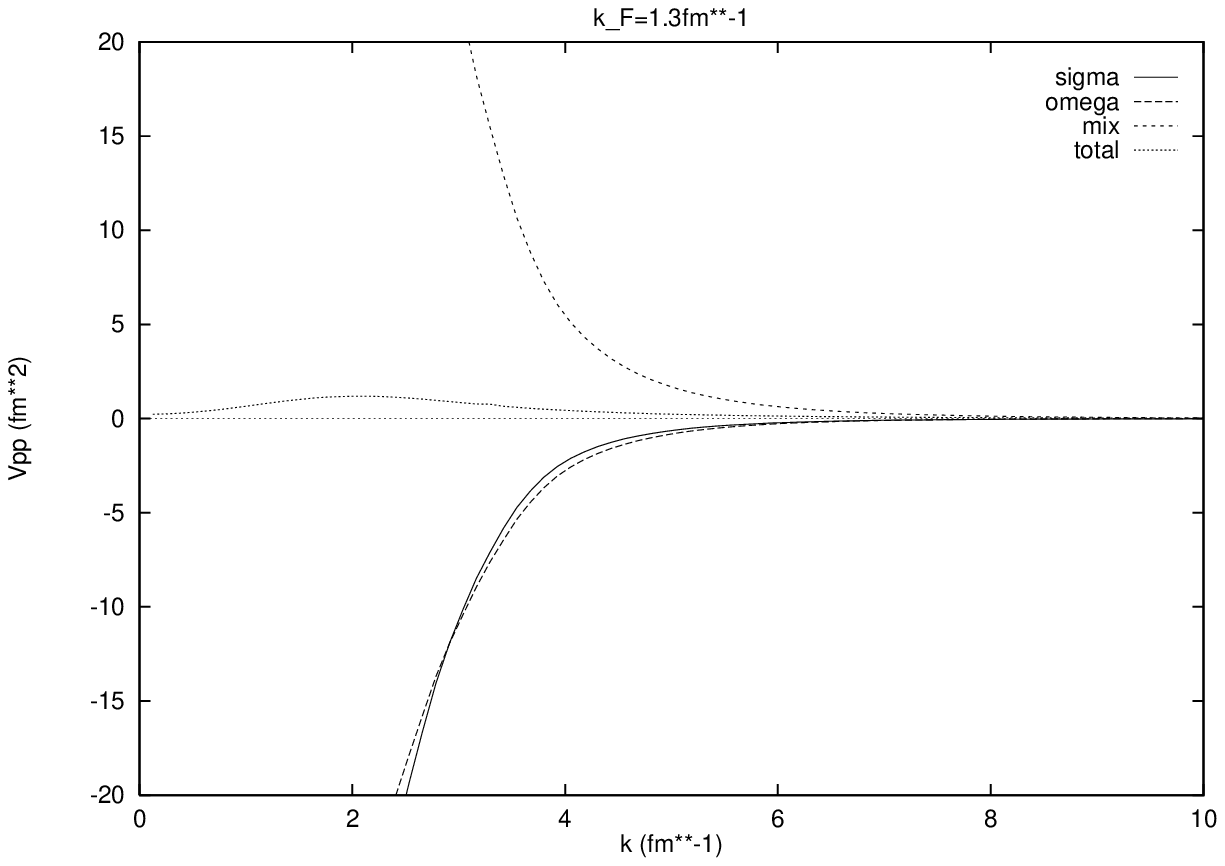,width=7cm}
\end{center}
\caption{The particle-particle interaction matrix element stemming from
the p-h polarization, $v_{\rm pp}^{({\rm pol})}(k,k_{\rm F})$, is shown as a
function of the momentum
$k$ for the Fermi momentum $k_{\rm F}$=0.7, 1.0 and 1.3 fm$^{-1}$ from top.
'total' is given by a strong cancellation between the attractions due to
$\sigma$ and due to $\omega$ and the repulsion due to the $\sigma$-$\omega$
mixing.}
\label{fig:figb}
\end{figure}

  Figure \ref{fig:figa} shows the result of the inclusion of the p-h
polarization.
The solid line indicates the pairing gap in neutron matter as a function of
the Fermi momentum given by the OBE interaction with the ordinary QHD-I
parameter set.  As mentioned earlier, the peak value is about three times
larger than the accepted values in the non-relativistic studies.  The dashed
line indicates the result given by the OBE + pure p-h polarization.  
This shows that the inclusion of the p-h polarization enhances the peak
value while the density range of the superfluid phase is reduced.  Contrary
to our expectation, this result is completely opposite to the non-relativistic
one shown in fig.5 in ref.\ref{refnine}, for example.  This is because
$v_{\rm pp}^{({\rm pol})}$,
which is given by a cancellation between the attractions due to $\sigma$ and
due to $\omega$ and the repulsion due to the $\sigma$-$\omega$ mixing, is
attractive at low densities while it is repulsive at high densities as shown
in fig.\ref{fig:figb}.
Further inclusion of the Pauli blocking, which is realized by adopting the
polarization insertion given by the density part of the Feynman-density
decomposition, enhances the gap especially at high densities as indicated by
the dotted line in fig.\ref{fig:figa}.

  Before drawing definite conclusions from the present result, two other
ingredients should be considered.  One is to include other mesons (such as
$\pi$ and $\rho$) although in the OBE level they were of no importance.  This
is suggested by an early non-relativistic study of finite nuclei that the
polarization due to the tensor force gave a strong repulsion in the $^1S_0$
channel~\cite{fifteen}.  The other
is to include the Dirac-sea polarization.  This is suggested by the work of
Guimar\~aes et al. which included the negative-energy states in the OBE
level~\cite{twelve}
and by the study of Friman and Henning that the $N$-$\bar N$ polarization
changed the sign of $\Pi(\vec{q})$~\cite{sixteen}, which will produce a sign
change of $v_{\rm pp}^{({\rm pol})}$ in the
present case.  Among these two ingredients which may influence the present
result, if the former is dominant the positive-energy sector of relativistic
models contains similar physics to non-relativistic models, whereas if the
latter is dominant the positive-energy sector alone cannot compare with
non-relativistic models.

  Finally, since QHD is an effective theory for hadronic many-body systems,
some high-momentum cut-off related to the nucleon size may be necessary.
Numerical results of the coupled eqs.(\ref{eq:eqa}) converge around
$\Lambda \simeq$10 fm$^{-1}$. The result presented in fig.\ref{fig:figa} was of
$\Lambda=$19 fm$^{-1}$;
this can be regarded as a full-range calculation.  We examined also some smaller
values of $\Lambda$.  As discussed in ref.\ref{refeleven}, it is possible to
reduce the absolute magnitude of the pairing gap by cutting off the
high-momentum contributions.  A prescription to determine $\Lambda$ was
proposed in ref.\ref{refthirteen}.  Since the p-h polarization is a
low-momentum process, the
qualitative feature that it enhances the peak value of the gap is not changed.

\section*{Acknowledgments}
  Discussions with Prof. R. Tamagaki, Prof. K. Matsuyanagi and Dr. H. Kouno are
acknowledged.

\end{document}